\documentstyle[12pt,amssym,aasms4]{article}

\newcommand{\secpoint}{\mbox{$''\mskip-7.6mu.\,$}}

\begin{document}
\title{The Low End of the Initial Mass Function \\ in Young  LMC Clusters: I. The Case of R136\altaffilmark{1}}
\author{Marco Sirianni\altaffilmark{2,3}, Antonella Nota\altaffilmark{3,4},
        Claus Leitherer\altaffilmark{3},
        Guido De Marchi\altaffilmark{5},  \\ and 
        Mark Clampin\altaffilmark{3}}
\altaffiltext{1}{Based on observations with the NASA/ESA {\it Hubble
Space Telescope}, obtained at the Space Telescope Science Institute,
which is operated by AURA for NASA under contract NAS5-26555, and
observations obtained at the European Southern Observatory, La Silla.}
\altaffiltext{2}{The Johns Hopkins University: sirianni@pha.jhu.edu}
\altaffiltext{3}{Space Telescope Science Institute, 3700 San Martin Drive,
Baltimore, MD 21218; nota@stsci.edu,
leitherer@stsci.edu, clampin@stsci.edu.}
\altaffiltext{4}{Affiliated with the Astrophysics Division, Space
Science Department of the European Space Agency.}
\altaffiltext{5}{European Southern Observatory: gdmarchi@eso.org}

\vspace{5cm}
\accepted{November 28,1999}
To appear in  {\it The Astrophysical Journal}\\

\newpage
\begin{abstract}
We report the result  of a study in which we have used very deep
broadband  V and I WFPC2 images of the  R136 cluster in the Large
Magellanic Cloud from the HST archive, to sample the luminosity
function below the detection limit of 2.8 M$_{\odot}$ previously
reached.  In these new deeper images, we detect  stars down to a
limiting magnitude of m$_{F555W}$ = 24.7 ($\simeq$ 1 magnitude deeper than previous
works), and identify a population of red stars evenly distributed  in
the surrounding of the R136 cluster.  A comparison of our
color-magnitude diagram with recentely computed evolutionary tracks
indicates that these red objects are  pre-main sequence stars in the
mass range  0.6 - 3 M$_{\odot}$.  We construct the initial mass
function (IMF) in the 1.35 - 6.5 M$_{\odot}$ range and find that, after correcting
for incompleteness, the IMF shows a definite flattening below  $\simeq$ 2
M$_{\odot}$.  We discuss the implications of this result for the R136
cluster and for  our understanding of  starburst galaxies formation and
evolution in general.
\end{abstract}
\newpage
\section{Introduction}\label{intro} 
The quest for a {\it universal} IMF has been a long standing issue
in stellar astrophysics (Scalo 1998). With the advent of the HST and
the improved sophistication of  ground based instrumentation, it has
been possible to extend to nearby galaxies studies that were in the
past feasible only in our own Milky Way, and at the same time reach the
new domain of the faintest and least massive stars, even before
they approach the main sequence. The studies of the IMF have expanded
in scope but also triggered new questions and added new
uncertainties.
 
For the field IMF, the uncertainties on distance and star
formation can be  so huge (Scalo 1998), that it would be difficult to
establish local variations. However, there is general agreement that
a  slightly steeper slope than Salpeter's ($\Gamma$ = -1.5 -- -1.8 {\it vs}
$\gamma$ = -1.35 in
the mass range 1 -- 10 M$_\odot$) is generally found.
 In the case of the  IMF for  star clusters and associations, the
distance effects are removed and  the star formation history is
simpler. For star clusters, at high masses (10 -- 100  M$_{\odot}$) there is good
agreement that the Salpeter IMF is ubiquitous. At small and intermediate
 masses (1 - 10 M$_{\odot}$) the situation is very different. Even in the LMC
itself, deep photometry of clusters has produced wildly discrepant
results, ranging from the very steep IMFs found by Mateo (1988)($\Gamma$
= -2.52), to the much  shallower slopes ($\Gamma$ $\simeq$ 0 - -1)
derived by Elson et al. (1989) and  Hunter et al. (1995, 1996).

30 Doradus in the LMC is the closest extragalactic HII region (Kennicutt
1991). It ideally offers a true laboratory for stellar population studies
because of its rich star formation history, and well determined  distance. It
is the local counterpart to distant starburst galaxies and  it has been
repeatedly defined as their {\it Rosetta Stone}.  However,  the emerging star
formation picture   is a diverse assembly of results which do not allow a
fruitful comparison, let alone an extrapolation to more distant galaxies.  The
situation becomes even more complicated when one considers independent studies
of the same region: for example, Oey and Massey (1995) derived $\Gamma$ = -1.3
$\pm$ 0.2 for the massive stars in the LMC superbubble LH47, while an
independent study by Will et al. (1997) quotes a resulting slope $\Gamma$ =
-2.1 in the same range of masses, using the same evolutionary models. However,
they then assumed a slope $\Gamma$ = -1.3 for the cluster over the entire mass
range investigated.   At the smallest masses, the discrepancies are even
larger: {\it are we observing true deviations from the Salpeter IMF}, most
likely triggered by local conditions of stellar density or star formation
history? Or are we simply dominated by the observational uncertainties,
related to the data analysis and interpretation, such as the choice of the
evolutionary models or treatment of completeness?

For all these reasons, we have started a systematic study of a number
of young clusters in the LMC and in the SMC, at different conditions of
stellar density, age and metallicity, with the objective of studying
the low end of the stellar IMF and to understand whether at small
masses the IMF is constrained by local conditions. The advantage of
such a  study is to reduce the uncertainties associated with data
reduction by establishing a homogeneous data treatment procedure
including a unique choice of models. This procedure will be applied to
all the clusters in the study, starting from the best known example,
R136, which we present in this paper. For this cluster, a number of
images exists in the HST archive which could be combined to produce deep
images of the cluster,  enabling us to  reach the low mass  limit of
0.6 M$_{\odot}$.

\section{Observations and data reduction}\label{obs}
Multiple images of the  R136 cluster were obtained with the WFPC2 on
board the HST in several bandpasses  after the first refurbishment
mission, as part of the Early Release Observations  and of the
WFPC2/GTO program. In particular, two sets of images were taken in
January 1994 and September 1994 as part of proposals 5589  and 5114,
which contained repeated
exposures in the filters F555W and F814W.  These filters are described
in detail by Biretta (1996) and closely resemble the Johnson V, I
filters in their photometric properties.  A journal of all observations
eventually combined is provided in Table~1.  All images were taken with
a gain of 7 e$^{-}$ ADU$^{-1}$. In both data sets, the R136 cluster was
centered in the Planetary Camera (PC), which has a field of view of
$35^{\prime\prime} \times 35^{\prime\prime}$,  with an effective plate
scale of $0\secpoint045$\,pixel$^{-1}$. The other three WF chips
observed flanking fields in 30 Doradus, with the same filter
configuration, but a larger field of view of 75$''$$\times$75$''$ per
chip and a plate scale of $0\secpoint1$\,pixel$^{-1}$.

The two datasets were processed independently using the standard STScI
pipeline procedure, which adopts standard  calibration  observations  and
reference  data, such as bias, flat field, and dark  frames  constantly
updated by  the WFPC2  team to track any changes in the performance of
the camera and its  detectors.  The basic  steps of the calibration
are the  correction  for the errors  introduced by the analog to
digital conversion, bias level and bias pixel-to-pixel variations
removal, dark image  subtraction,   flat  field  image application
and  shutter  shading corrections.

The images taken in January and September, however, were characterized by 
different temperatures of the CCDs, and, therefore, different
charge transfer characteristics. In fact, in March 1994 a significant
charge transfer efficiency (CTE) variation had been found in WFPC2,
which caused a 10-15 \% gradient in  the photometric response of the
CCDs along the columns of each chip.  This  effect is due to the
partial loss of signal when charge is transferred down the chip during
the readout, with the consequence that stars at higher row numbers
appear fainter than they would if they were at low row numbers
(Holtzman et al. 1995a). A significant reduction to the   CTE effect
was achieved by cooling down the four CCDs  from -76 to -88 $^o$C. The new
temperature became operational on April 23 1994, and the CTE stabilised
at a  $\sim$ 4 \% level. In order to account for the difference in CTE
between the two data sets, different corrections were performed: a 12\%
correction ramp was applied to the  -76$^o$C data (January 1994), and a
4\% correction was applied to the  -88$^o$ C data (September 1994) in
order  to bring the charge packets of each pixel to the values they
would have had in the absence of the CTE problem.

The images were then registered and combined to remove cosmic rays.  A rotation
of 99.85$^{o}$ was applied to the final image from the first set, which was also
shifted in the horizontal and vertical direction by an amount (+74.07, +134.73).
The combined images,  670 $\times$ 590 pixel in size, have an overall exposure
time of 1240 sec and 760 sec in the filters F555W and F814W respectively.  It
should be pointed out that due to the presence of detector read noise, the total 
combined exposure time is not  fully equivalent to the same time in a  single
image, and will eventually yield a slightly lower S/N.  The combined  Planetary
Camera image for the filter F555W is shown in Figure~1,  with the R136 cluster in
the center.

\section{The photometry}\label{phot}
Photometry has been performed on both images, using the PSF fitting
routines provided within DAOPHOT. The first step in the photometric
reduction procedure was to discriminate  between {\it true} stars and
spurios objects which might have been introduced  by both the alignment
procedure and the hot pixels removal. This was done by studying in
detail the characteristics of the stellar PSF. We carefully selected by
eye a sample of  150 {\it bona fide} stars in the  final F555W image.
A statistical study of this sample allowed us to define an appropriate
range for the parameters that DAOFIND uses as selection criteria.  Two
parameters are particularly important: the {\it roundness}, which
allows  us  to eliminate objects which are too elongated along rows or
columns,  and the {\it sharpness}, which eliminates objects whose
profile differs largely from a gaussian profile. From our sample of 150
stars we found mean values of $0.78 \pm 0.1$ and $0.045 \pm 0.165$ for
the sharpness and roundness, respectively.

We then ran DAOFIND on out data, by conservatively setting the detection
threshold at $4\,\sigma$ above the local background level, but excluded any
object with sharpness and roundness parameters exceeding by more than  $\pm
3\,\sigma$ the average bona fide values (Figure~2). We inspected the
rejected objects and found that almost all of them were  noise peaks associated
with hot columns, diffraction spikes or highly  saturated stars, with a small
number being isolated hot pixels and  extended objects. Due to the extreme
saturation of the central regions,  the innermost $2^{\prime\prime}$ of the
cluster core were excluded from  our study.

The list of stars detected in the F555W combined image was then used to identify
the stars in the F814W image: 1706  stars were found to be common to both frames. 
In order to carry out the  photometry with the highest accuracy possible, we first
performed aperture photometry and  measured the stellar flux in a very small
aperture (2 pixel radius), selected to match the FWHM of the PSF ($\simeq$ 2
pixel). We measured the background as the mode of the annulus centered  on each
star with a inner radius  of 3 and an outer radius of 7 pixels. We  then
constructed a sample PSF by combining three moderately bright and isolated stars,
located as close as possible to the  unsaturated central region of the frame.   In
order to properly subtract the background  it was necessary to carefully evaluate
the aperture correction, accounting for  the fraction of source  light  present in
the background annulus. Such a correction is derived by   assessing how the PSF
encircled energy varies as a function of the  distance from the peak.  For each
image we have measured the encircled  energy profile for a number of isolated
stars and have used these  measurements to correct the fluxes derived with
aperture photometry.

  The procedure of rotation and shift of the two dataset  just
marginally modified  the final PSF FWHM from the original values of
1.51 pixel and 1.32 pixel the  for the two  separate datasets to 1.76
pixel in the final F555W combined frame.  Although this procedure
resulted in a  slight degradation of the spatial resolution, photometry
tests clearly demonstrate this to be  much preferable to the alternative of
performing photometry on the individual images.

The magnitudes of the individual stars were  then determined relative to an
aperture of radius $0\secpoint5$, and transformations were made to translate the
on-orbit system to the WFPC2 photometric system using the calibration provided by
Holtzman et al. (1995b).  The zeropoints used were $22.48$ mag for the F555W
filter and $21.60$ mag for the F814W band.  The original zeropoints were provided
for a gain of 14 e$^{-}$ ADU$^{-1}$ to which we have added the correction for the
different gain adopted in these observations (7 e$^{-}$ ADU$^{-1}$) (Holtzman et
al. 1995b).

In Figure~3 we report the photometric errors assigned by DAOPHOT to all our
measurements: for our study, we discarded all stars with an associated error
larger than $ 0.2$ magnitudes in both filters.  With this limitation, we find
1604 stars common to  both filters, down to a limiting magnitude m$_{\rm F555W}$ 
= 24.7, which is $\sim$1 magnitude deeper than the individual frames published by
Hunter et al. (1995, 1996). Altough we do not include the full photometry in this
paper, the complete table is available in electronic form upon request.

\section{The Color-Magnitude diagram}

We generated an observed color-magnitude diagram (CMD) where m$_{\rm F555W}$ is
plotted as a function of the (m$_{\rm F555W}$ - m$_{\rm F814W}$) color 
(Figure~4), which includes  all the stars measured in the combined F555W and
F814W frames with photometric errors smaller than  0.2 magnitudes.  The CMD
immediately reveals  the presence of two distinct branches, segregated in color
at  (m$_{\rm F555W}$ - m$_{\rm F814W}$)  $\simeq$ 1.1.

This color segregation is  very similar to the  effect  observed   in the CMD of 
NGC 3603, the galactic clone of R136 (Drissen, 1999). The most likely origin of
this effect could be the presence of differential reddening within the cluster,
or to the {\it real}  presence of a second population of faint redder stars.

\subsection{Differential reddening in R136?}

The distribution of gas and dust in the region surrounding R136 is highly
inhomogeneous, as can easily be seen in images of the region taken in the light
of H$\alpha$, [OIII] and [SII] (Scowen et al. 1998).  Ideally, the most
accurate strategy would be to redetermine the reddening coefficients for each
individual star, and indeed Hunter et al. (1995, 1996) had attempted to do so
in their recent papers. Unfortunately, the uncertainties associated with their
data, as well as the red leak present in the UV filter they adopted, made this
task unsuccessful, and they opted instead for the use of ground based
measureaments by Fitzpatrick \& Savage (1984).  They  eventually attributed the
spread in color they observed in their CMD to the use of a single coefficient.

Adopting the same reddening law of Hunter et al. (1995), originally derived by
Fitzpatrick \& Savage (1984), we have: E(B-V) = 0.38 and R$_{v}$ = 3.4. 
Converted to our filters, this yields A$_{\rm F555W}$ = 1.37,  and A$_{\rm
F814W}$ = 0.80. In Figure~4  we have
superimposed  such reddening vector on the observed CMD. \\ There is consensus that differential
reddening exists in the R136 region. However, if differential reddening were the
cause for the bimodal distribution  observed in our CMD, the total  amount  of
absorption  needed would be very high.  We should in fact  assume  an extreme
value of E(B-V) = 0.51, and R$_{v}$ = 6.08, to match  the two observed features.
This value, which is reported in Figure~4 for comparison, is  more than three
times higher than the observed values, and totally inconsistent with  the ground
based measurements. We  are  therefore confident to exclude the possibility of
such a high differential extinction and conclude that the observed split in the
CMD is  most likely intrinsic, and due to the presence of a second population of
fainter, redder stars.

\subsection{The red population: evidence for pre-main sequence stars?}

The position of the red stars  in the CMD  is consistent with a
population of pre-main sequence (PMS) stars, as already suspected by Hunter
et al. (1995).  In order to establish whether this scenario is correct,
we needed to construct isochrones  to assign masses to the stars and
determine ages for the stellar population. We used isochrones
constructed from the  stellar evolution models provided by Siess et
al.  (1997) for PMS evolution.  Siess's tracks offer the choice of
different metallicities: Z = 0.02, 0.04, 0.005.  We adopted Z = 0.005 for
this study, which closely matches the LMC values of Z = 0.008. 

The stellar models are provided in the [log(L/L$_{\odot}$) {\it vs}
log($T_{\rm eff}$)] plane.  In order to convert the information to the
WFPC2 [(m$_{\rm F555W}$)$_o$ {\it vs}(m$_{\rm F555W}$)$_o$ - (m$_{\rm
F814W}$)$_o$] system, we used the model atmospheres of  Kurucz (1993)
interpolated at  metallicity  Z = 0.005  and g = 4.0  and the standard STSDAS SYNPHOT software
package to reproduce the WFPC2 photometric response.  For each
isochrone,  we assigned to each point in the plane ($T_{\rm eff}$, log
g)  the corresponding Kurucz's atmospheric model and evaluated the
intrinsic color (m$_{\rm F555W}$ - m$_{\rm F814W}$)$_{o}$ of such a
point using SYNPHOT.  We  then converted log(L/L$_{\odot}$) into the
(m$_{\rm F555W}$)$_{o}$ magnitude adopting a distance modulus of 18.6
(Walborn et al. 1997), and the Bolometric Correction (BC) provided by
Bessel et al. (1998) for the same conditions of gravity.

In Figure~5 we show the dereddened CMD of the  R136 cluster, where (m$_{\rm
F555W}$)$_{o}$ is plotted as a function of (m$_{\rm F555W}$ - m$_{\rm
F814W}$)$_{o}$.  We have superimposed to the CMD the isochrones corresponding
to the PMS evolutionary tracks in the range 
5 $\times 10^5  - 5  \times 10^7$ yr. As expected,  the observed red population is
very well tracked by these PMS isochrones, and  most likely consists of low
mass stars (down to 0.6 M$_\odot$) still approaching the main sequence.  Their
age is found to be in the range between  1 and 10 Myr. In Figure~5 we also
show, for comparison, the position of a star of 1.35, 1.5 and 2  M$_\odot$ on
the various isochrones.

\subsection{The age of the R136 cluster and the pre-main sequence
stars} 

There is consensus in the published work that the mean age of the
R136 cluster is less than $5 \times 10^{6}$ yr.  Already at
the time when the nature of the central object in R136  was still
unclear, Savage et al. (1983) and Schmidt-Kaler \& Feitzinger (1982)
had proposed an age of 2 $\times$ 10$^{6}$ yr  for the {\it
supermassive} central object.  With the discovery of WR features in the
integrated spectrum of the central region, the age determination
increased (Melnick 1985).  Later on, Campbell et al.  (1992) took high resolution
HST/WFPC images of the cluster core, and argued that the presence of
WR stars in the central region suggested an age of at least 3.5
$\times$ 10$^{6}$ yr.  Almost at the same time, De Marchi et al. (1993)
used the WR stars to place both a lower and upper limit to the age; in
fact, while the simple presence of WR stars argues in favour of an age
higher than 2  $\times$ 10$^{6}$ yr, the fact that the WR stars are of
the WNL type, a hydrogen rich subclass which is usually associated with
very massive progenitors ($>$ 50 M$_{\odot}$), sets  a firm upper limit
of 5 $\times$ 10$^{6}$ yr.  Such an evidence, combined  with the lack
of red supergiants  found by Campbell et al. (1992), led De Marchi et
al.  (1993) to conclude in favour of an age determination of 3 $\times$
10$^{6}$ yr for the R136 cluster. The most recent work by de~Koter et al.
(1998) suggests that the WR stars in R136 are younger than classical
WNL stars. If so, R136 has an age of at most 2~Myr, and may even be
somewhat younger.

In the cluster, star formation is almost coeval: the R136 cluster
core extends over a linear scale $< 10$\,pc.  Consider the typical
time scale associated with the star formation process ($t = d/v$, where $v$
is the propagation speed of the shock wave triggering star formation).
Typical oberserved values of $v$ are of order 50 km s${-1}$ (e.g., Satyapal
et al. 1997), so that the possible age spread is less than $0.5$\,Myr.
 This is small compared to the evolutionary timescale of the stars
formed (De Marchi et al. 1993).  Outside the R136 cluster, star formation
is most likely still continuing (Walborn 1984), especially in the outer 
filaments of the 30 Doradus region. 

If we assume coeval star formation and an age for the cluster of $\simeq 2 - 4
\times 10^{6}$\,yr, we find that   all stars down to 1\,M$_{\odot}$ have
already reached their birthline, defined as the locus in the HR diagram along
which young stars first appear as visible objects (Stahler 1983).  In fact,
stars of $1 - 2$\,M$_{\odot}$ reach their birthline in less than $0.5$\,Myr.

As already mentioned, we find that the population of young stars in the
R136 CMD  is well fitted by isochrones of  10$^{6}$  --  5 $\times$ 10$^{7}$ yr,
indicating that we are observing  the stars  while they are approaching
the ZAMS.  Typically it  takes $\simeq$  5 $\times$ 10$^{7}$ yr for a star of 1.5
M$_{\odot}$ to reach the ZAMS (Siess et al. 1997). This interval is
shorter for stars of higher mass. A star of 2 M$_{\odot}$ will take
$\simeq$ 3 $\times$ 10$^{7}$ yr, and a star of 3 M$_{\odot}$ about
 $\simeq$ 1 $\times$ 10$^{7}$ yr. It is  safe to conclude that the red extension of the R136
CMD is made of PMS objects in the 1-3 M$_{\odot}$ range,
observed in their approach to the ZAMS. 

Did all these stars  form at the same epoch of the R136 cluster? As
already pointed out by Hunter et al. (1995),  stars down to 3
M$_{\odot}$ appear to have formed approximately at the same time of the
more massive stars.   However, the  least massive stars (1 -- 2 M$_{\odot}$)
could possibly be a few  Myr older, but the uncertainties associated
with both the theoretical isochrones and the observational data are
such that, at this point, no precise answer can be  provided.

\section{The H-R diagram} 
In an attempt to determine the IMF of R136, we need  to locate the stars
onto the H-R diagram (HRD). We do so by  converting the
photometric information (magnitude and color) into the (Log T$_{\rm
eff}$ {\it vs} Log L/L$_\odot$) plane.

To this purpose, we have used the relation (V - I) {\it
vs}  $T_{\rm eff}$, for g = 4.0, from Bessel et al. (1998). This relation is
suitable for PMS stars as well as  normal stars (see Sung et al. 1998). From
the same  source, we also assume BC as a function of $T_{\rm eff}$. 
In order to use these relations, we have converted m$_{\rm F555W}$ and  m$_{\rm F814W}$ into V and I using the equations provided by Holtzmann et al. (1995b).

We  then adopted  a distance modulus of $(m-M)_o$ = 18.6 (Walborn et al.
1997) to translate the apparent magnitudes into absolute luminosity, as
follows:
\begin{center} 
$Log(L/L_\odot) = 0.4 \times (4.75 - M_V - BC)$ \\
\end{center}

where M$_{V}$ = M$_{F555W}$ - C, being C a correction factor derived
for each star from the Holtzmann et al. (1995b).  Figure~6 shows
the theoretical HRD with the superimposed  evolutionary tracks for the mass range
0.6 -- 7 M$_\odot$  from Siess et al. (1997). \\ The HRD further illustrates the
composition of the intermediate-low mass stellar population in R136 and
surroundings: stars with mass above 4 M$_{\odot}$ are already on the main
sequence or in close proximity.  Stars at lower masses (0.6 -- 3.0 M$_{\odot}$)
display a higher concentration in proximity to their birth line (at the redward
origin of their evolutionary tracks in Figure~6) and have not reached the ZAMS
yet. It is interesting to notice that while stars above 3 M$_{\odot}$ evolve at
almost constant luminosity  to the ZAMS, at smaller masses stars do  experience
quite significant variations in effective temperature and  luminosity, thus
creating a quite large observed spread in both quantities. The gap observed  at
approximately T$_{eff}$ $\simeq$ 3.8 -- 3.9 between the two populations reflects
an evolutionary effect.  The evolution  from the birth line is faster for high
mass stars:  stars of 4 -- 7 M$_{\odot}$ will transition in that  T$_{eff}$
region much faster that the smallest stars, and therefore they will be observed
in smaller numbers, thus creating the observed  opening. At the smallest masses
(1 -- 1.5 M$_{\odot}$), the gap is less noticeable.

\subsection{The completeness and the photometric errors}

Before proceeding to the derivation of the IMF it is also necessary to
establish the completeness of our data.  With this objective in mind, we have
defined four regions  (A, B, C, D; see Figure~7) surrounding the R136
cluster  in the final
combined F555W image.  As  can be noticed in Figure~7, these four
regions have different characteristics in terms of gas/dust contamination and
crowding. Region A is the most crowded, and includes many very bright stars,
while region C displays some obscurations due to dust and gas.  Regions B and D
are intermediate in their properties.  For each region,  and for each filter,
the completeness has been assessed with the following procedure:
\begin{itemize}
\item The sample of stars falling within the region has been divided into
fifteen half magnitude bins;
\item Artificial stars have been added to each magnitude bin, in quantity
not to exceed 10\% of the total number, in order not to affect severely
the crowding in the region considered. Numerous tests have been run with
the same recipe (100 per bin);
\item The artificial stars have been retrieved, using the same
selection criteria of sharpness and  roundness adopted in our work.
Stars with photometric error larger than 0.2 magnitudes have been
discarded.
\end{itemize}
The results of the test are summarized in Table~\ref{compll}, where for each
region (A -- D) and filter, we have reported the completess factor as a
percentage of the stars  successfully retrieved {\it vs} the total number of
stars artificially added.  As  can be noticed in Table~\ref{compll}, the
completeness is worse  in the brightest magnitude bin, 
where saturation effects prevents the detection of other very bright objects,
and  towards
the faint end, where S/N effects start to dominate. In region A, which is
characterized by many bright stars, the completeness  drops significantly  much
earlier than for the other regions.  For regions B, C, and D  the agreeement is
quite good  and the completeness is quite robust (better than 50 \%) down to
m$_{F555W}$ = 23.2, m$_{F814W}$ = 22.1.  We  have used an average of regions B, C,
and D  to derive  completeness correction factors for our photometry. We 
have  used the completeness factors determined in this way to draw {\it
completeness lines} onto the HRD to underline the variation of the correction
factors with luminosity (Figure~8).  As it can be seen in the  figure,
the completeness is very robust down to Log (L/L$_{\odot}$) $\simeq$ 0.5. 
In order to understand how this result
impacts the definition of a conservative lower mass limit to our measurements,
we have constructed in  Figure~9, for each of the four regions
considered - A,B,C, and D - a completeness histogram as a function of the
corresponding mass.   As already  discussed, regions B, C, and D are in quite
good  agreement, while region A displays the largest deviation. We have
therefore assumed that regions B, C, D are homogeneus in their properties, and
we show in  Figure~9 their average completeness (solid
line). For the average of regions B, C and D  we find that the completeness
correction drops below 50\%  at $\sim$ 1.35 M$_\odot$.  This will be
the conservative lowest mass limit to our measurements. \\

Having established the completeness of our photometry, it was necessary to 
estimate how our photometric errors  affect the  location of the stars in the
HRD and, therefore, the determination of their mass. To this purpose, we have
considered four different regions in the CMD, which are representative of
different luminosities and temperatures, and sample the range of values present
in our CMD. For each region, we have averaged magnitude and colors for ten
stars, in order to derive a single representative point, with a mean magnitude
and color. For this representative point, we also derived  a mean error, in
magnitude and color, as an indication of the uncertainties associated with
stars in that region of the CMD.  We  then transformed these four
representative points, and their associated errors, into the HRD
(Figure~10), and  found that these typical errors on the photometry
translate into a small error in luminosity and  into a larger
error in T$_{eff}$.  However, since all evolutionary tracks down to 3
M$_{\odot}$  develop at  almost constant T$_{eff}$, we are confident that such
an error in the temperature does not lead to a significant uncertainty in the
mass determination, at least for masses higher than 3 M$_{\odot}$.  For
masses below 3 M$_{\odot}$, a translated error of 0.2 in  Log T$_{eff}$ can
affect the mass determination by shifting the star to the adjacent mass bin.
This effect increases towards  smaller masses, where we should assume an
worse case uncertainty on the mass determination of $\pm$ 0.3 M$_{\odot}$.  

\section{The Initial Mass Function}
\subsection{The Initial Mass Function of the R136 cluster}
The IMF, usually indicated  by $\xi$, is
defined as the number of stars per logarithmic mass interval per unit
area. The slope of the IMF is given by $\Gamma = d(log \xi)/d(log M)$
where the standard IMF (Salpeter 1955) has a slope  $\Gamma= -1.35$.
Although we use the term IMF, we are actually discussing the present day
mass function (MF). In the case of the very
young R136, where star formation has been coeval, we can safely assume
that the observed MF is the IMF.

The MF of stellar clusters is usually measured by counting the number
of stars as a function of the magnitude (luminosity function) which is
then converted into the number of stars per unit stellar mass by the
use of the appropriate mass-luminosity relation. This approach,
however, can only be applied to stars currently on their main sequence,
i.e. to objects for which a precise correspondence exists between mass
and luminosity. As Figure~4 shows, however, R136 hosts a large
population of PMS objects, for which such a relation depends
strongly on the age and is, therefore, very difficult to apply.

An alternative avenue to follow, which has the advantage of overcoming
this age degeneracy, is that presented by Tarrab (1982) and based on
the use of the HRD and theoretical isomass tracks in place of the CMD.
In order to derive the MF of R136, we have counted the number of stars
falling between each pair of  tracks shown in the HR diagram
(Figure~6) and normalized such number  to the width of the mass
range spanned by the tracks and to the area of the observed field. As
already mentioned, we have adopted the evolutionary tracks by Siess et
al.(1997), which also include  PMS stars down to 0.6 M$_\odot$. 
The number of stars in each mass bin is shown in Table~\ref{imftab}
(column 3), together with the width of the bin (column 2). To properly
account for the effects of crowding, we have corrected the numbers
measured in this way for the incompleteness of our photometry, using
the {\it completeness factors}  described above (see Table~\ref{compll}).
The values corrected in this way are also listed in Table~\ref{imftab}
(column 4). 

The determination of the MF has been carried out independently in the four
regions A, B, C, and D.  Since these regions are so differently affected by
crowding  and dust/gas contamination, and are associated with different
completeness corrections, the comparison of their MFs  provides an
independent assessment of the solidity of our results. Also, we have have
limited our  MF measurements to the magnitude range where the completeness is
robust, that is better than 50\%.   For the average of regions B, C and D the
completeness correction drops below 50\%  at $\sim$ 1.35 M$_\odot$
(see Figure~9). This is the conservative lower mass limit we have
assumed for our MF determination.  

We find that, within the limitations imposed by small numbers statistics,
there is good agreement among the four MFs derived in this way.
Following a conservative approach, we have then proceeded to discard
the  most extreme region (A) (see Figure~9)  and have retained only B, C and D
for the construction of the final MF, which is shown in Figure~11, 
where the  errors on the data points account for both the Poisson 
statistics of the counting process and for the uncertainty on the 
completeness.

Two  different trends are distinguishable in the IMF profile of
Figure~11: for stars in the mass range $2.1 - 6.5$\,M$_{\odot}$ the data
points are well fit by a slope with $\Gamma$ = -1.28 $\pm$ 0.05, while at lower
masses the IMF profile flattens out, with  a derived slope 
$\Gamma$ = -0.27 $\pm$ 0.08 (1.35 -- 2.1 M$_{\odot}$). For comparison, the
Salpeter slope is provided on the edge of the figure.

\subsection{The Mass Function of the surrounding areas of R136}

For comparison, we obtained the MF of three flanking fields, which were
observed by the three WF chips while the R136 cluster was centered in
the field of view of the Planetary Camera. These  WF fields covered
each a region of $\simeq 75^{\prime\prime} \times 75^{\prime\prime}$,
and together a total area of $\simeq$  4.7 arcmin$^{2}$. 
 At a distance for the LMC of 52.5  Kpc (Walborn et al. 1997), this
corresponds to an area of 1092 pc$^{2}$. For reference, we
provide the central location of all WF chips in Table~\ref{wf}.  Since
the January and September datasets were taken with different
orientations, it was not possible to add the two datasets and reach the
same depth as in the combined PC images. Only the September dataset was
considered for this work: we reached a  magnitude limit in the summed
F555W frame of m$_{\rm F555W} = 24$ with a combined exposure time of 840\,s.
A total of 2836 stars were found in both filters, selected with the
same conservative criterion described above to discard any object with
an associated photometric error larger than 0.2 magnitudes in both
filters.

A  CMD was constructed in similar fashion to what was done for the
R136 cluster, and is  shown in Figure~12. The most relevant difference
from  the R136 cluster CMD is a better defined main sequence and  the total
absence of the second  population of red stars. Although the depth of the 
flanking field images differs from that of the  R136 cluster,
we can exclude the presence of a second redder population, segregated in color. 
This can be  seen from the insert of Figure~12, which
shows the histogram of the number of objects in function of the color (m$_{\rm
F555W}$ - m$_{\rm F814W}$) for  the R136  cluster (solid line) and the flanking
fields (dashed line).  The two distributions appear completely different.

  The MF was derived following the same procedure described for R136, and
the final result is shown in Figure~13. In this case, the   filled
circles represent our observational data, which are also listed in
Table~\ref{fimftab} together with the completeness correction factors derived
adopting the same procedure.  Again, we have used all data points with a photometric completion
better than 50\%.  The MF slope derived in this way is slightly shallower than
that of Salpeter, with $\Gamma = -1.23 \pm 0.11$. 

Can we directly compare the derived  MF to the R136 IMF?  Naturally, when
considering field stars,   the additional uncertainty on the distance of the
stars considered, and their age, has to be accounted for. Therefore, a direct
quantitative comparison cannot be made.

\subsection{The spatial distribution of the  pre-main sequence stars}
 A first visual inspection  of the images indicates that PMS objects
are  quite uniformly distributed within the field of the Planetary Camera. In
order to define more accurately their distribution, we divided the entire 
field into concentric annuli centered on  R136a  up to a distance of 13$''$ from
the center of the cluster.  As already mentioned, we discarded the inner
$2^{\prime\prime}$ radius region, since the  cluster core  is highly
saturated.  We  defined each annulus to be  50 pixels wide, corresponding to
$2\secpoint25$.

We  then compared the  spatial distribution,  averaged within each
annulus, of the PMS stars (m$_{F555W}$ - m$_{F814W}$ $ > $ 1.1,
M $<$ 3 M$_{\odot}$)
and of the more massive stars  (m$_{F555W}$ - m$_{F814W}$ $< $ 1.1, 
M $>$ 3 M$_{\odot}$).  The
two distributions are illustrated in Figure~14, where the
number of stars per surface area is provided as a function of distance
from the cluster center. The solid line  with the star symbols 
indicates the PMS
stars and the solid lines with the filled circles the more massive ones. \\
 Both distributions have been corrected for incompleteness. To this
purpose, we have performed   a second   test  to establish the
completeness level within each annulus, following the same procedure
previously described. The results of this test are listed in
Table~\ref{cdist}, in terms of completeness correction factors for the
various annuli. Due to the high number of saturated stars in the
central region, we  find that the first two annuli (up to 4.5$''$ from
the center) are severely affected by incompleteness, and therefore no
conclusions can be drawn in the inner cluster regions. \\
 At a distance of 5$''$ from the center and further out, the two
distributions are comparable, although the most massive stars display a
somewhat steeper decrease in number as we move from the center towards
the outer regions of the cluster. The population of pre-main sequence
stars appears uniformly distributed  in the  distance range
 5 - 13$''$, with a very slight increase towards the central regions.

\section{On the universality of the IMF}

The IMF of R136 has been recently studied by Hunter et al. (1995,
1996). Their determination agrees well with ours down to $\sim
3$\,M$_\odot$, where their data are reliable. Below this limit,
however, their errors are so large to make the IMF determination
tremendously uncertain and, as such, not meaningful. In all cases, the
shape of the IMF at masses larger than $\sim 3$\,M$_\odot$ is
compatible with a power law with index $\Gamma \simeq
-1.2$, extending up to $\sim 6.5$\,M$_\odot$ in our study and all the way
to $\sim 15$\,M$_\odot$ in theirs. The agreement with the canonical IMF
of Salpeter (1955) is thus preserved (see Figure\,11). Our determination of the IMF is consistent with the work
of Sagar \& Richtler (1991), who have studied the intermediate mass
range in five LMC clusters finding an average slope $\Gamma \simeq
-1.1$ in the range $2 - 12$\,M$_\odot$.

At lower masses, however, our data are clearly different,
 indicating a flattening or possibly a drop  below $\sim
2$\,M$_\odot$ in the logarithmic plane. This does not mean that the
number of objects is no longer increasing with decreasing mass, rather
that the increase proceeds at a lower pace. Although in principle
crowding could be at the origin of this effect, the flattening of the
IMF occurs where our photometry is robust, with a completeness better 
than $\sim 50$\,\%.  

A similar effect, i.e. a deficiency of stars in the $1 - 2$\,M$_\odot$
range, is also observed by Hillenbrand (1997) in the Orion Nebula
Cluster, although in that case the plateau is followed first by a steep
increase between $0.5$\,M$_\odot$ and $0.2$\,M$_\odot$ and then by a
clear drop all the way down to the H-burning limit. Although there are
several examples of a flat IMF for stars less massive than $\sim
1$\,M$_\odot$ (see e.g. Comeron, Rieke, \& Rieke (1996) in NGC\,2024
and Scalo (1998) for a review of the IMF in the solar neighborhood),
only in $\rho$\,Oph have Williams et al. (1995) found a flat IMF {\it
above} 1 M$_{\odot}$. The question as to whether the flattening that we
observe in R136 is characteristic of this cluster or a general feature
cannot therefore be conclusively addressed, and our finding might
simply add on to the conclusion of Scalo (1998) that, at least in this
mass range, the IMF is far from being uniform in the Universe.

\section{Conclusions}

The main results of this study are:

\begin{itemize}
\item We have detected stars with masses as low as 0.6~M$_{\odot}$ in the R136 cluster
using archival HST-WFPC2 images. The least massive stars in our study are about 
1~mag fainter than stars known from previous work. 
\item The lowest mass  stars in R136 are identified as a population of
pre-main-sequence stars from a comparison with Siess et al.  (1997) evolutionary
models.
\item By combining the pre-main-sequence and the main-sequence population, we
are able to derive the stellar IMF between  1.35 and 6.5~M$_{\odot}$. The
IMF in this mass range does no longer follow a power law but begins flattening
below  $\sim$2~M$_{\odot}$.
\end{itemize}

Should we call the low-mass IMF in R136 peculiar? It would certainly be too
simple-minded to assume that the power-law IMF observed in the high-mass range
could extend  all the way to lower and lower masses. Evidence for a flattening of
the IMF at the low-mass end is manifold in the solar neighborhood (Scalo 1998).
Yet, this  flattening does not normally set in at masses of $\sim$2~M$_{\odot}$.
Observations of both field stars and clusters in the solar neighborhood suggest a
flattening around 0.3~M$_{\odot}$, an order of magnitude lower than in R136.
Almost all Galactic clusters for which the low-mass IMF is known are different
from R136 in their stellar content: they contain few, if any, massive stars with
masses above 10~M$_{\odot}$. R136, in contrast, has now been shown to contain
$\sim$10$^3$ O stars {\em and} a significant low-mass population.

The Galactic massive-star formation regions NGC3603 and NGC6231 both
have IMF determinations based on star counts. Eisenhauer et al. (1998)
find no evidence for an IMF flattening in NGC3603 down to about
1~M$_{\odot}$. Sung, Bessell, \& Lee (1998) on the other hand report a
clear deficit of stars with masses below 2.5~M$_{\odot}$ in NGC6231,
the center of the Sco~OB1 association. At the higher masses, the IMF in
NGC6231 is close to Salpeter. R136 and NGC6231 appear to have rather
similar IMF over the mass range 1 to 100~M$_{\odot}$.

Although low-mass stars are clearly forming in R136 (and other Galactic
regions of massive-star formation), they do not form with the same
frequency as more massive stars. This is reminiscent of the IMF in
starburst galaxies, for which a deficit of low-mass stars has been
suggested (Rieke 1991). The low-mass end of the starburst IMF is not
accessible to direct observations but must be inferred dynamically.
Therefore uncertainties are large and alternative interpretations have
been proposed (e.g., Satyapal et al. 1997). The low-mass end of the
R136 IMF is not completely dissimilar to an IMF truncated at a few
solar masses. The total masses in stars following our derived IMF is
very close to the mass obtained from a power-law IMF with a Salpeter
slope ($\Gamma=-1.35$) above 1~M$_{\odot}$ and no stars below that mass.
Although not quite as extreme as the ``top-heavy'' starburst IMF, the
R136 IMF may indicate a real difference in the mass spectrum of stars
formed in- and outside starbursts, possibly related
to the gas density of the interstellar medium.

\acknowledgements

\clearpage
\begin{figure}
\figurenum{1}
\caption{Final combined WFPC2 image of the R136 cluster in the F555W filter.
The image shows the portion of the  field of view of the Planetary Camera (30$''$
$\times$ 27$''$), where the cluster is centered. The orientation on the sky
is indicated in the image.}
\label{f555w}
\end{figure}

\begin{figure}
\figurenum{2}
\caption{A sample of {\it bona fide} stars selected by eye has been used to statistically
derive the appropriate values for the "sharpness" and "roundness" parameters
used by DAOFIND to discern the {\it true } stars from spurious artifacts.}
\label{daofind}
\end{figure}
\begin{figure}
\figurenum{3}
\caption{Photometric errors assigned by DAOPHOT to all our measurements. For our
study, we have discarded all stars with an associated error larger than
0.2 magnitudes in both filters.
}
\label{errors}
\end{figure}
\begin{figure}
\figurenum{4}
\caption{Observed Color Magnitude Diagram of R136, for all stars measured in the combined  
images with associated photometric error smaller than 0.2 in both filters.
We have
superimposed the reddening vector  originally derived by Fitzpatrick \& Savage
(1984), where E(B-V) = 0.38 and R$_{v}$ = 3.4.  In order to explain with
differential reddening the bimodal distribution  observed in our CMD, the
total  amount  of absorption  needed would be  E(B-V) = 0.51, and R$_{v}$ = 6.08. 
This value is also  reported in the figure for
comparison. }
\label{cmdr}
\end{figure}
\begin{figure}
\figurenum{5}
\caption{Dereddened Color Magnitude Diagram of the R136 cluster, to which we have
superimposed isochrones corresponding to $5 \times 10^{5}  -  5 \times 10^{7}$ yr. The
{\it red population} is well fitted by pre-main sequence isochrones,
and most likely consists of low mass stars still approaching the main
sequence. We also show, for comparison, the position of a star of 1.35, 1.5
 and 2 M$_{\odot}$ on the various isochrones.}
\label{cmdis}
\end{figure}
\begin{figure}
\figurenum{6}
\caption{H-R diagram for all the stars observed with photometric errors lower
than 0.2 magnitudes. We have superimposed to the HRD the
pre-main sequence evolutionary tracks for the mass range 0.6 - 7 M$_{\odot}$ 
from Siess et al. (1997). The dashed lines indicate  post MS evolution. }
\label{hr}
\end{figure}
\begin{figure}
\figurenum{7}
\caption{The final combined image in the F555W filter has been subdivided in
four regions surrounding the R136 cluster, with the objective of
establishing the completeness of our photometry. The four regions
display different characteristics in terms of crowding and/or gas/dust
contamination. For each region we indicate nomenclature and size (in $''$.)
}
\label{compl}
\end{figure}
\begin{figure}
\figurenum{8}
\caption{H-R diagram for all the stars observed with photometric errors lower
than 0.2 magnitudes. We have superimposed to the HRD the {\it completeness
lines} derived from the execution of the completeness test executed
on regions B,C and D.}
\label{hrc}
\end{figure}
\begin{figure}
\figurenum{9}
\caption{A completeness histogram as a function of mass for the four selected
regions around the R136 cluster: A, B, C, and D. The average completeness for
regions B, C and D, which appear to display homegeneous properties, is drawn as
a solid line. The mass corresponding to the conservative limit of 50\% for the
completeness correction factor has been chosen as the lowest limit to our MF
determination. This corresponds to M = 1.35 M$_\odot$. }
\label{ABCD}
\end{figure}

\begin{figure}
\figurenum{10}
\caption{H-R diagram for all the stars observed, to which we have
superimposed the error boxes corresponding to  four representative stars
(and their associated  photometric  errors) selected in the CMD. This figure
illustrates how the photometric errors propagate in the transformation
to the HRD.  While the resulting error in the luminosity is small,
the photometric error translates into a larger error in T$_{eff}$.}
\label{hrer}
\end{figure}
\begin{figure}
\figurenum{11}
\caption{The IMF for the R136 cluster, obtained by averaging the
IMFs for the regions B, C, and D, defined as the number of stars per unit
logarithmic mass per square parsec. The  full circles indicate the 
completeness-corrected IMF, while the  the open diamonds  and the stars
 indicate the slopes from Hunter et al. (1995, 1996). A Salpeter slope is provided
for reference.}
\label{imf}
\end{figure}
\begin{figure}
\figurenum{12}
\caption{Color Magnitude Diagram for the combination of  three fields flanking
the R136 cluster, down to a limiting magnitude m$_{\rm F555W}$ = 24. The stars included have  associated photometric error smaller than 0.2 magnitudes in both filters.}
\label{fcmd}
\end{figure}
\begin{figure}
\figurenum{13}
\caption{The IMF for the R136 cluster flanking fields, obtained 
from the three adjacent WF chips. The total area investigated is
$\simeq$ 4.7 arcmin$^{2}$, with a limiting magnitude m$_{F555W}$ = 24.}
\label{fimf}
\end{figure}
\begin{figure}
\figurenum{14}
\caption{Spatial distribution for the pre-main sequence stars as a function
of distance with respect to the R136 cluster center  (crosses) compared
with the distribution of the more massive stars (full dots).}
\label{dist}
\end{figure}

\clearpage
\singlespace
\begin{table}
\caption{ Journal of {\it HST$+$WFPC2} Observations \label{journal}}
\begin{tabular}{lllcl}
\tableline
\tableline
Proposal & Date (UT) & Filter &  Exposure Time (sec) & Image Name\\
\tableline
5589  &   January 1994    &  F555W  & 200 & U25Y0109T \\
5589  &   January 1994    &  F555W  & 200 & U25Y0101T \\
5589  &   January 1994    &  F814W  & 100 & U25Y0207T \\
5589  &   January 1994    &  F814W  & 100 & U25Y0208T \\
\tableline
5114  &   September 1994  &  F555W  & 120 & U2HK030JT \\
5114  &   September 1994  &  F555W  & 120 & U2HK030KT \\
5114  &   September 1994  &  F555W  & 120 & U2HK030LT \\
5114  &   September 1994  &  F555W  & 120 & U2HK030MT \\
5114  &   September 1994  &  F555W  & 120 & U2HK030NT \\
5114  &   September 1994  &  F555W  & 120 & U2HK030OT \\
5114  &   September 1994  &  F555W  & 120 & U2HK030PT \\
5114  &   September 1994  &  F814W  &  80 & U2HK0317T \\
5114  &   September 1994  &  F814W  &  80 & U2HK0318T \\
5114  &   September 1994  &  F814W  &  80 & U2HK0319T \\
5114  &   September 1994  &  F814W  &  80 & U2HK031AT \\
5114  &   September 1994  &  F814W  &  80 & U2HK031BT \\
5114  &   September 1994  &  F814W  &  80 & U2HK031CT \\
5114  &   September 1994  &  F814W  &  80 & U2HK031DT \\
\tableline
\tableline
\end{tabular}
\\
\end{table}

\begin{table}
\caption{Completeness correction factor for regions A, B, C and D\label{compll}}
\begin{center}
\begin{tabular}{ccccc||ccccc}
\hline 
\hline
m$_{F555W}$ &    A  &   B   &   C  &     D   &   m$_{F814W}$ &           A  &   B   &   C  &     D   \\
\hline                                                                                            
17.7  &         64  &   68  &   67  &   70   & 16.1  &                87 &    90 &    89 &    91  \\   
18.2  &         84  &   88  &   90  &   89   & 16.6  &                89 &    89 &    91 &    93  \\   
18.7  &         85  &   85  &   87  &   89   & 17.1  &                89 &    91 &    92 &    93  \\   
19.2  &         80  &   83  &   86  &   89   & 17.6  &                85 &    87 &    90 &    90  \\   
19.7  &         79  &   81  &   85  &   87   & 18.1  &                85 &    87 &    88 &    90  \\   
20.2  &         69  &   76  &   81  &   84   & 18.6  &                79 &    84 &    85 &    88  \\   
20.7  &         67  &   75  &   78  &   83   & 19.1  &                74 &    81 &    83 &    85  \\   
21.2  &         61  &   71  &   73  &   83   & 19.6  &                70 &    72 &    79 &    85  \\   
21.7  &         54  &   67  &   70  &   80   & 20.1  &                64 &    75 &    73 &    83  \\   
22.2  &         44  &   57  &   61  &   78   & 20.6  &                57 &    65 &    71 &    81  \\   
22.7  &         32  &   50  &   60  &   69   & 21.1  &                45 &    62 &    68 &    71  \\   
23.2  &         25  &   43  &   55  &   63   & 21.6  &                41 &    53 &    61 &    71  \\   
23.7  &         14  &   37  &   42  &   48   & 22.1  &                25 &    45 &    52 &    61  \\   
24.2  &         8   &   19  &   19  &   22   & 22.6  &                11 &    17 &    18 &    19  \\   
24.7  &         0   &   3   &   1   &   2    & 23.1  &                2  &    6  &    8  &    8   \\   

\hline
\end{tabular}
\end{center} 
\end{table}  

\begin{table}
\caption{The IMF for the R136 cluster\label{imftab}}
\begin{tabular}{cccc}
\hline 
\hline
Mass (M$_{\odot}$) &    Bin width (M$_{\odot}$) & \# stars/bin  & corrected \# stars/bin  \\
\hline

     0.650000  &   0.100000   &    14   &    64 \\
     0.750000  &   0.100000   &    12  &    181 \\
     0.850000  &   0.100000   &    20   &    66 \\
     0.950000  &   0.100000   &    11   &    31 \\
      1.05000  &   0.100000   &    10   &    76 \\
      1.15000  &   0.100000   &    15   &    98 \\
      1.25000  &   0.100000   &    28   &    254 \\
      1.35000  &   0.100000   &    30   &    70 \\
      1.45000  &   0.100000   &    43   &    78 \\
      1.55000  &   0.100000   &    37   &    64 \\
      1.65000  &   0.100000   &    39   &    60 \\
      1.75000  &   0.100000   &    34   &    48 \\
      1.85000  &   0.100000   &    40   &    57 \\
      1.95000  &   0.100000   &    38   &    51 \\
      2.10000  &   0.200000   &    63   &    84 \\
      2.35000  &   0.300000   &    78   &    102 \\
      2.75000  &   0.500000   &    91   &    116 \\
      3.50000  &    1.00000   &    92   &    111 \\
      4.50000  &    1.00000   &    58   &    68 \\
      5.50000  &    1.00000   &    44   &    53 \\
      6.50000  &    1.00000   &    20   &    26 \\
\hline
\end{tabular}
\end{table}

\begin{table}
\caption{Coordinates of the  WFPC2 flanking field centers\label{wf}}
\begin{tabular}{cll}
\tableline
\tableline
CHIP     &     R.A.         &   DEC \\
WF2      & 5:38:49.27       &  -69:05:17.50 \\
WF3      & 5:38:57.23       &  -69:06:12.72 \\
WF4      & 5:38:46.68       &  -69:06:56.70  \\
\tableline
\tableline
\end{tabular}
\end{table}

\begin{table}
\caption{The IMF for the Field\label{fimftab}}
\begin{tabular}{cccc}
\hline 
\hline
Mass (M$_{\odot}$) &    Bin width (M$_{\odot}$) & \# stars/bin  & corrected \# stars/bin  \\
\hline
      1.05000  &   0.100000   &    2   &    5 \\
      1.15000  &   0.100000   &    14   &    41  \\
      1.25000  &   0.100000   &    63   &    626   \\
      1.35000  &   0.100000   &    179   &    484 \\
      1.45000  &   0.100000   &    156   &    214 \\
      1.55000  &   0.100000   &    163   &    195 \\
      1.65000  &   0.100000   &    176   &    201 \\
      1.75000  &   0.100000   &    129   &    145 \\
      1.85000  &   0.100000   &    126   &    143 \\
      1.95000  &   0.100000   &    104   &    111 \\
      2.10000  &   0.200000   &    179   &    189 \\
      2.35000  &   0.300000   &    168   &    176 \\
      2.75000  &   0.500000   &    258   &    267 \\
      3.50000  &    1.00000   &    331   &    341 \\
      4.50000  &    1.00000   &    217   &    222 \\
      5.50000  &    1.00000   &    108   &    110 \\
      6.50000  &    1.00000   &    85   &    87 \\

\hline
\end{tabular}
\end{table}

\begin{table}
\caption{Spatial Distribution completeness test\label{cdist}}
\begin{tabular}{c|ccccc}
\hline 
\hline
M$_{F555W}$ &   \multicolumn{5}{c}{Distance}  \\
                   & 3.375' & 5.625'' & 7.875'' & 10.125 '' & 12.375'' \\
\hline

17.7  &         58  &   64  &   64  &   69   &   73    \\   
18.2  &         77  &   86  &   87  &   90   &   88    \\   
18.7  &         71  &   84  &   86  &   90   &   88    \\   
19.2  &         67  &   80  &   81  &   86   &   88    \\   
19.7  &         56  &   74  &   79  &   84   &   80    \\   
20.2  &         50  &   69  &   72  &   81   &   77    \\   
20.7  &         40  &   62  &   71  &   78   &   74    \\   
21.2  &         31  &   56  &   66  &   70   &   67    \\   
21.7  &         25  &   48  &   61  &   66   &   62  \\   
22.2  &         18  &   42  &   53  &   57   &   59  \\   
22.7  &         9  &   33  &   43  &   51   &    52 \\   
23.2  &         2  &   26  &   35  &   42   &    44    \\   
23.7  &         0  &   18  &   28  &   32   &    32    \\   
24.2  &         0   &   7  &   13  &   14   &    15    \\   
24.7  &         0   &   1   &   2   &   3   &     2   \\

\hline
\end{tabular}
\end{table}

\clearpage
\clearpage
\clearpage
\plotone{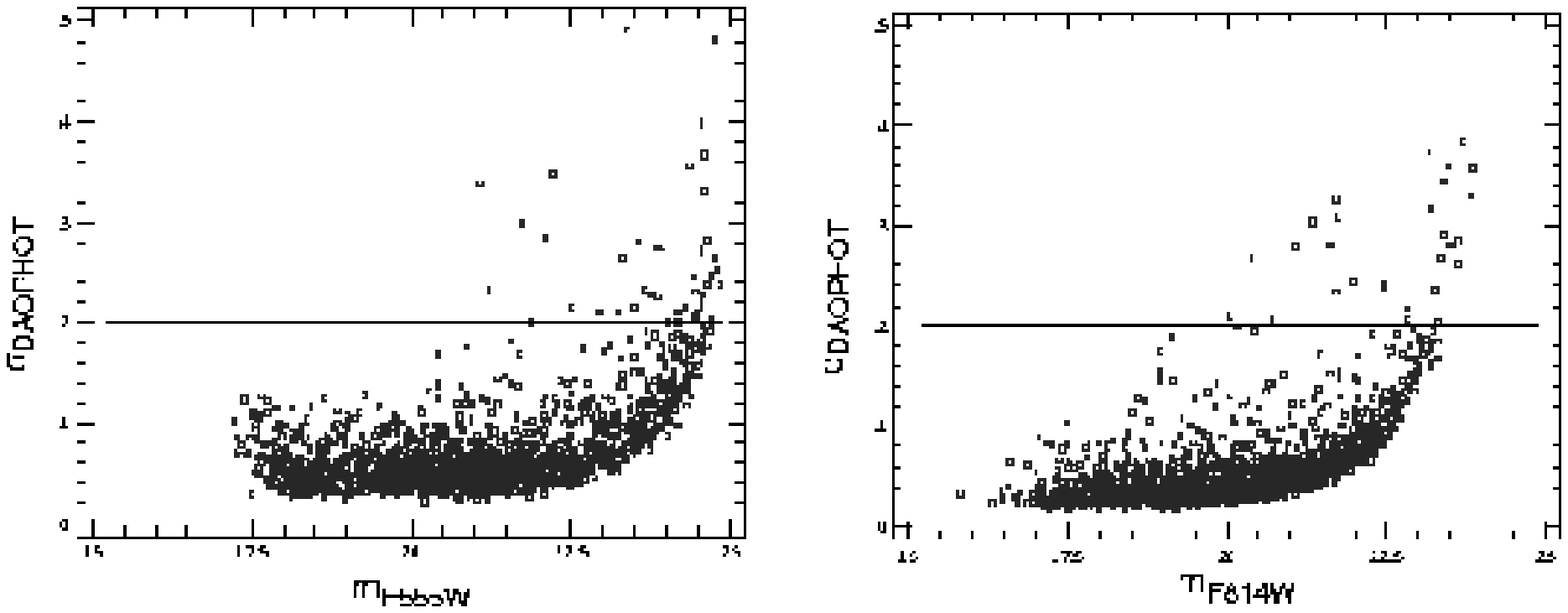}
\clearpage
\plotone{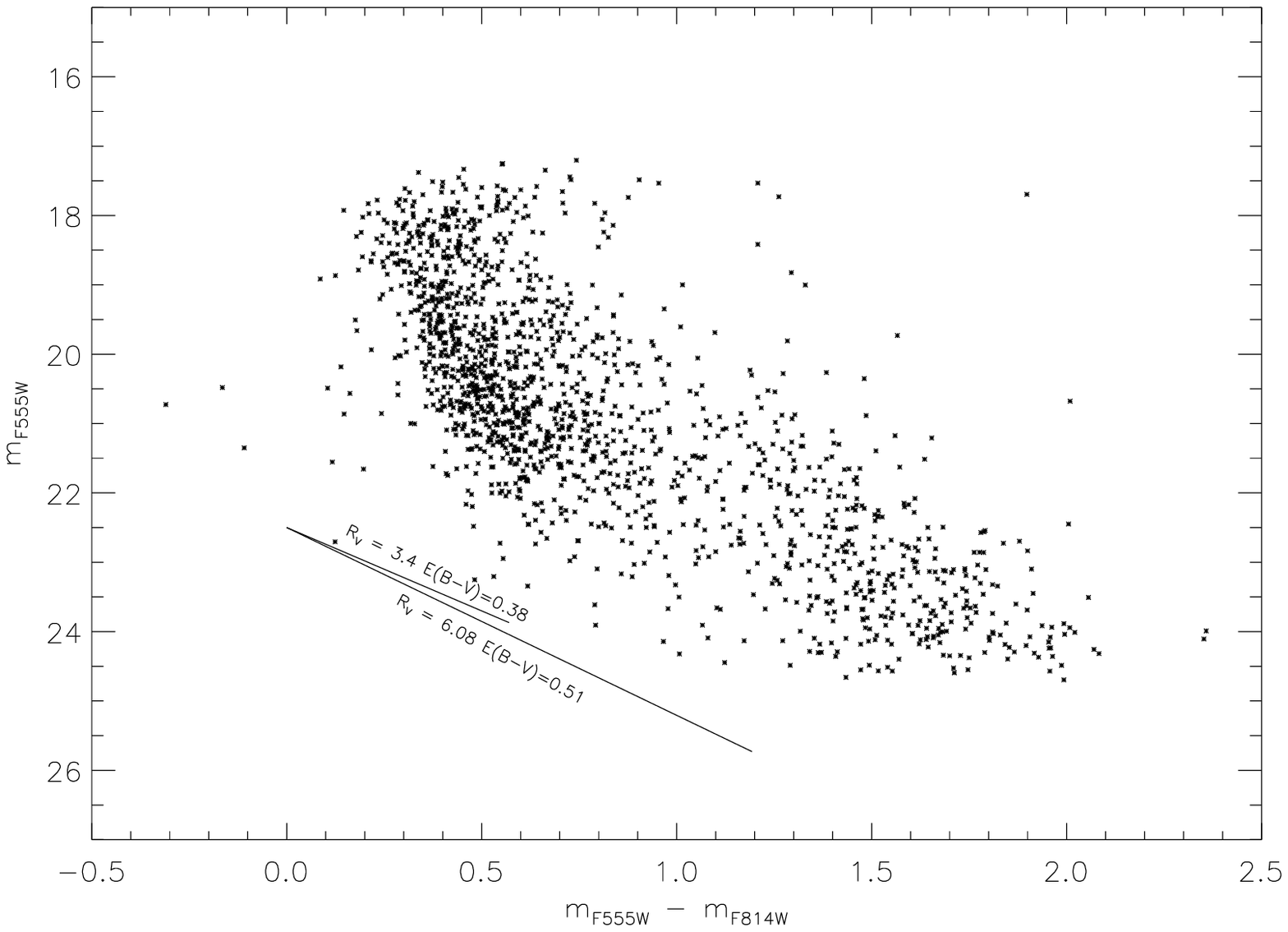}
\clearpage
\plotone{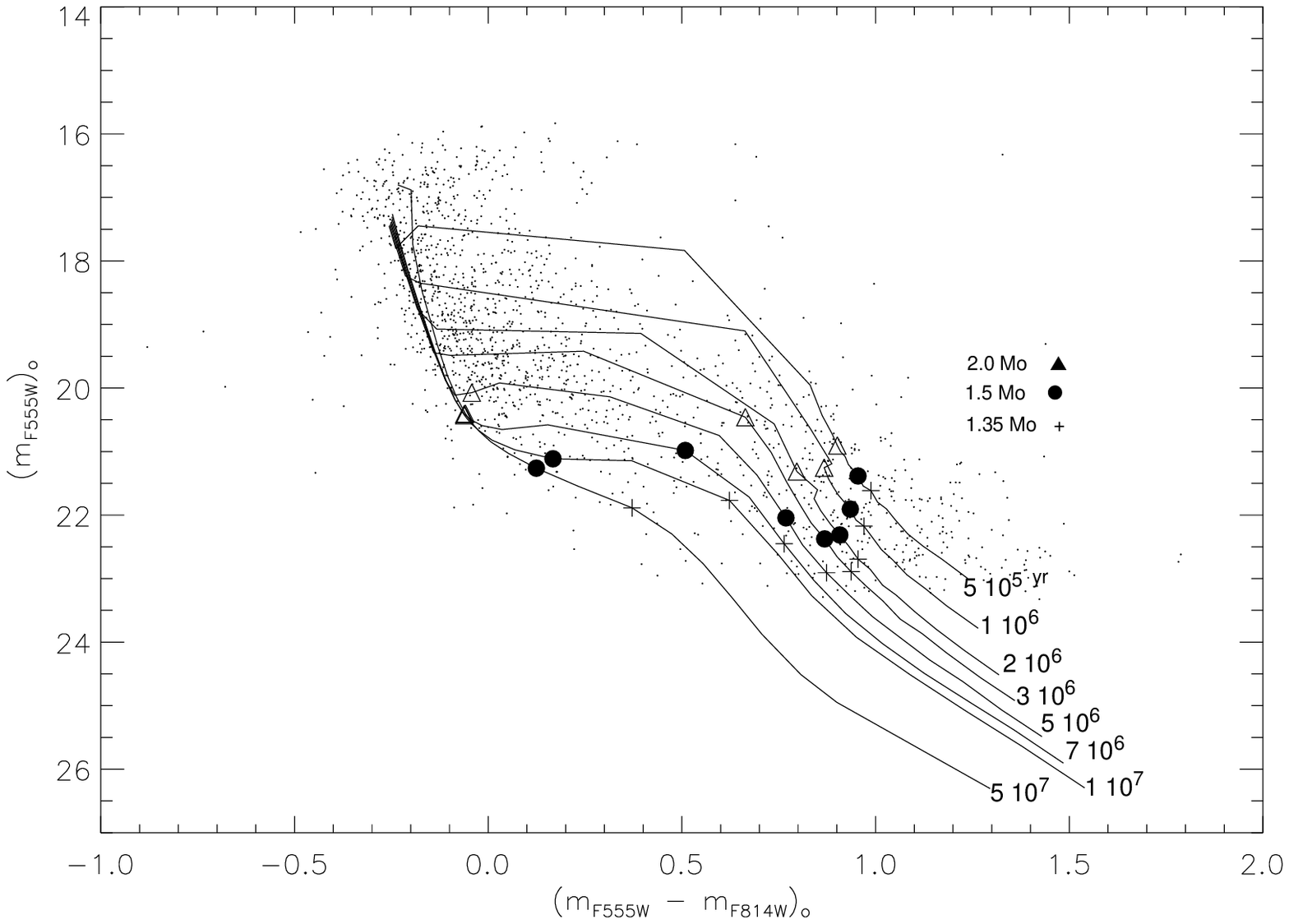}
\clearpage
\plotone{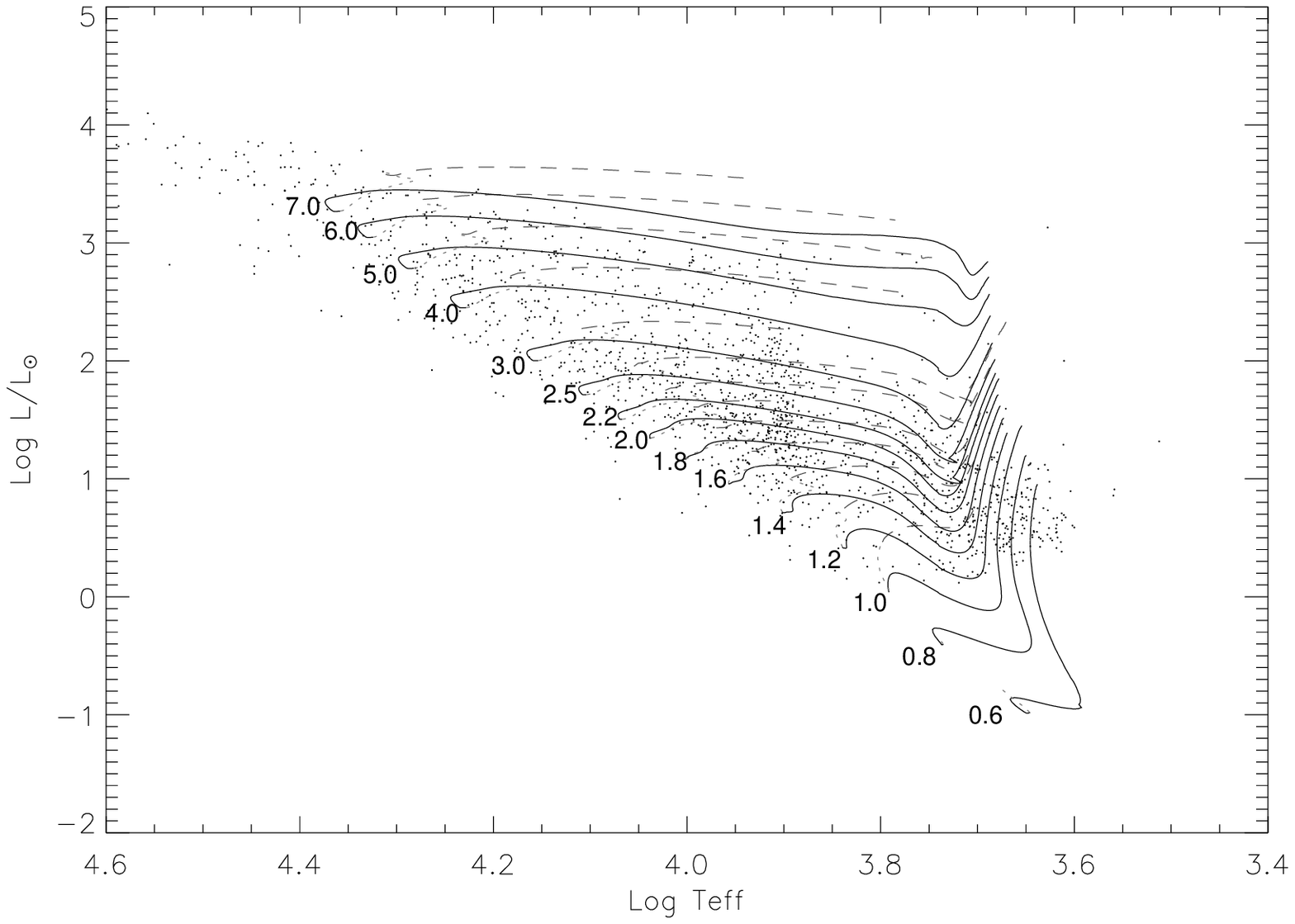}
\clearpage
\clearpage
\plotone{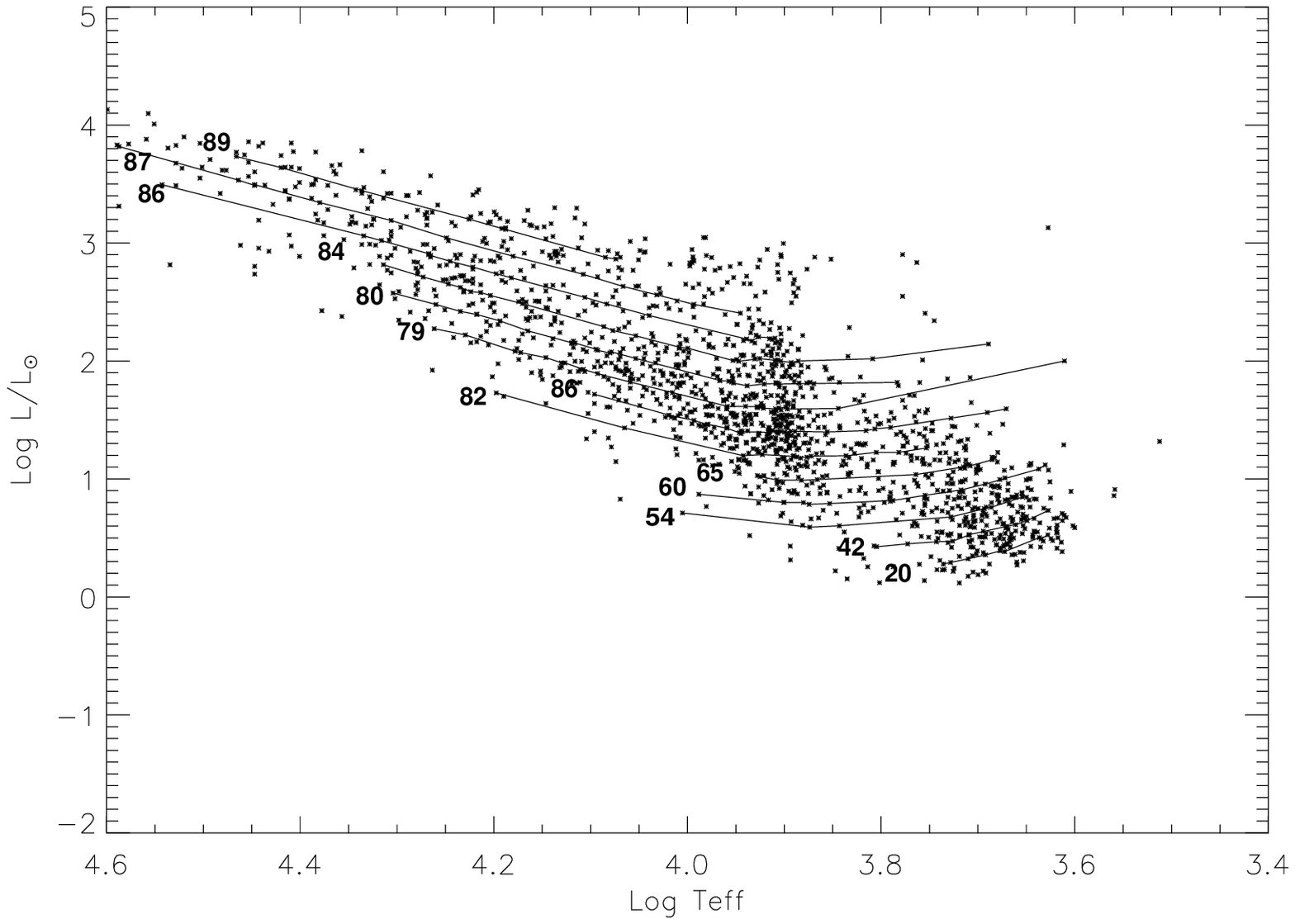}
\clearpage
\plotone{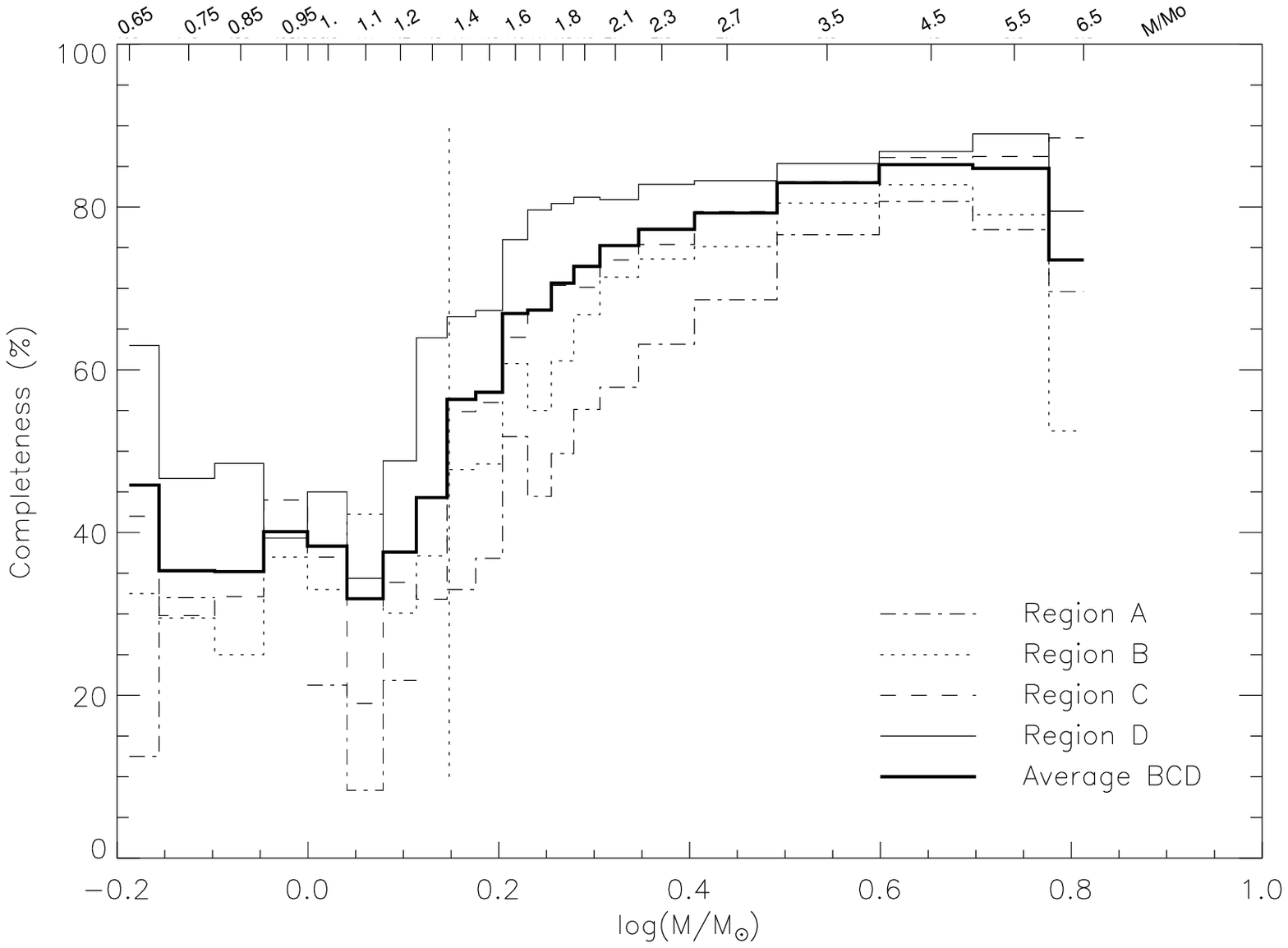}
\clearpage
\plotone{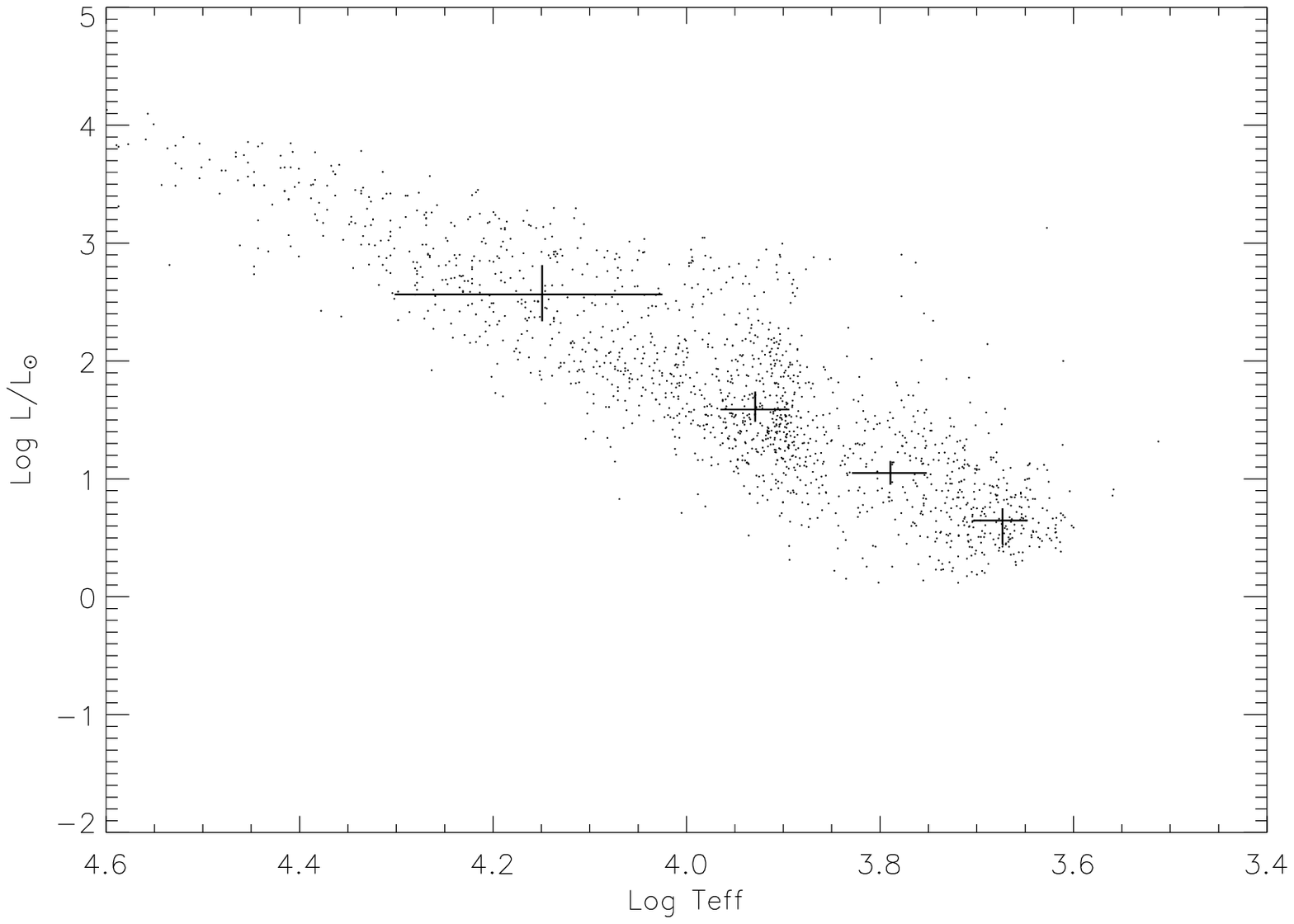}
\clearpage
\plotone{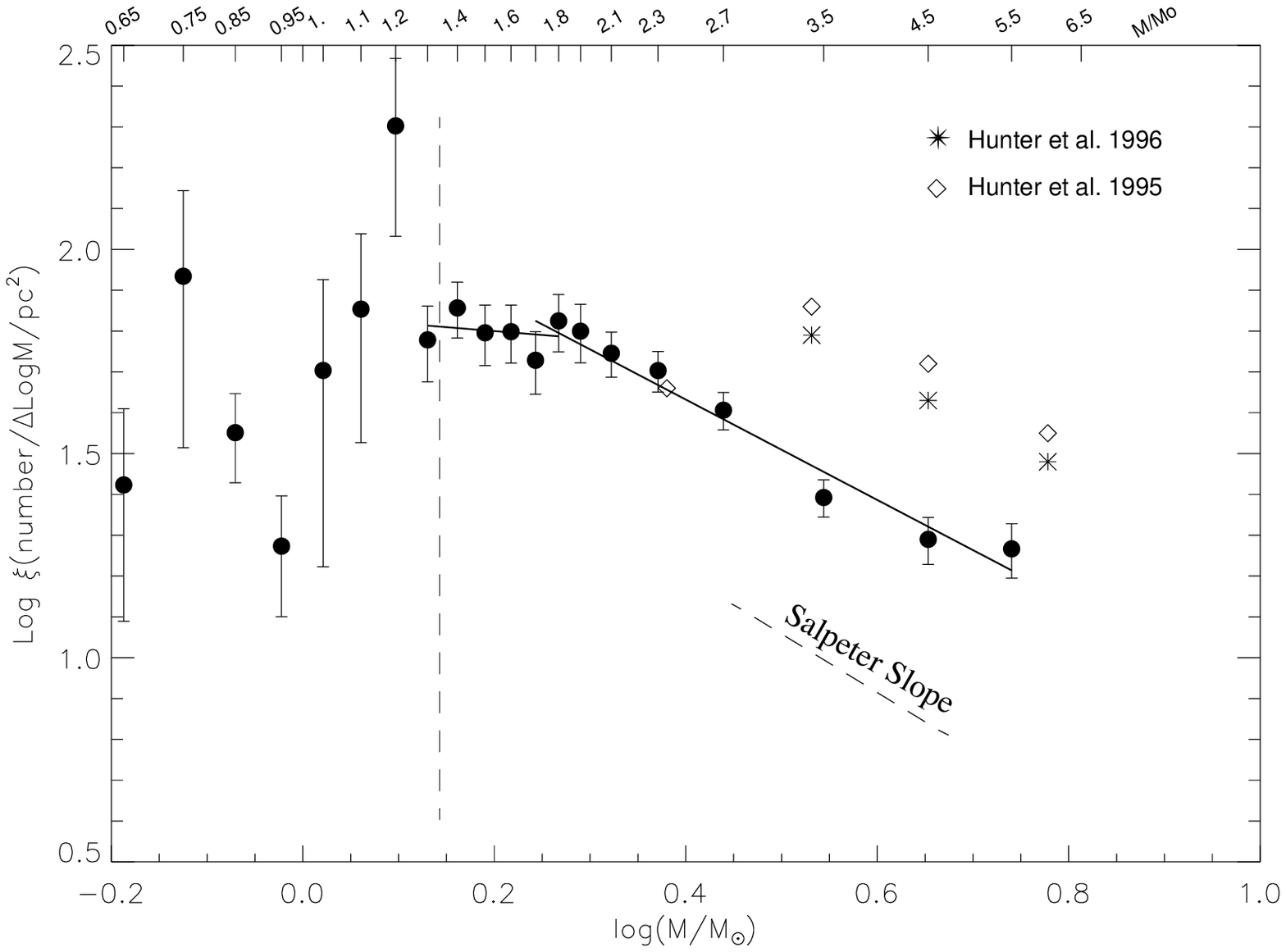}
\clearpage
\plotone{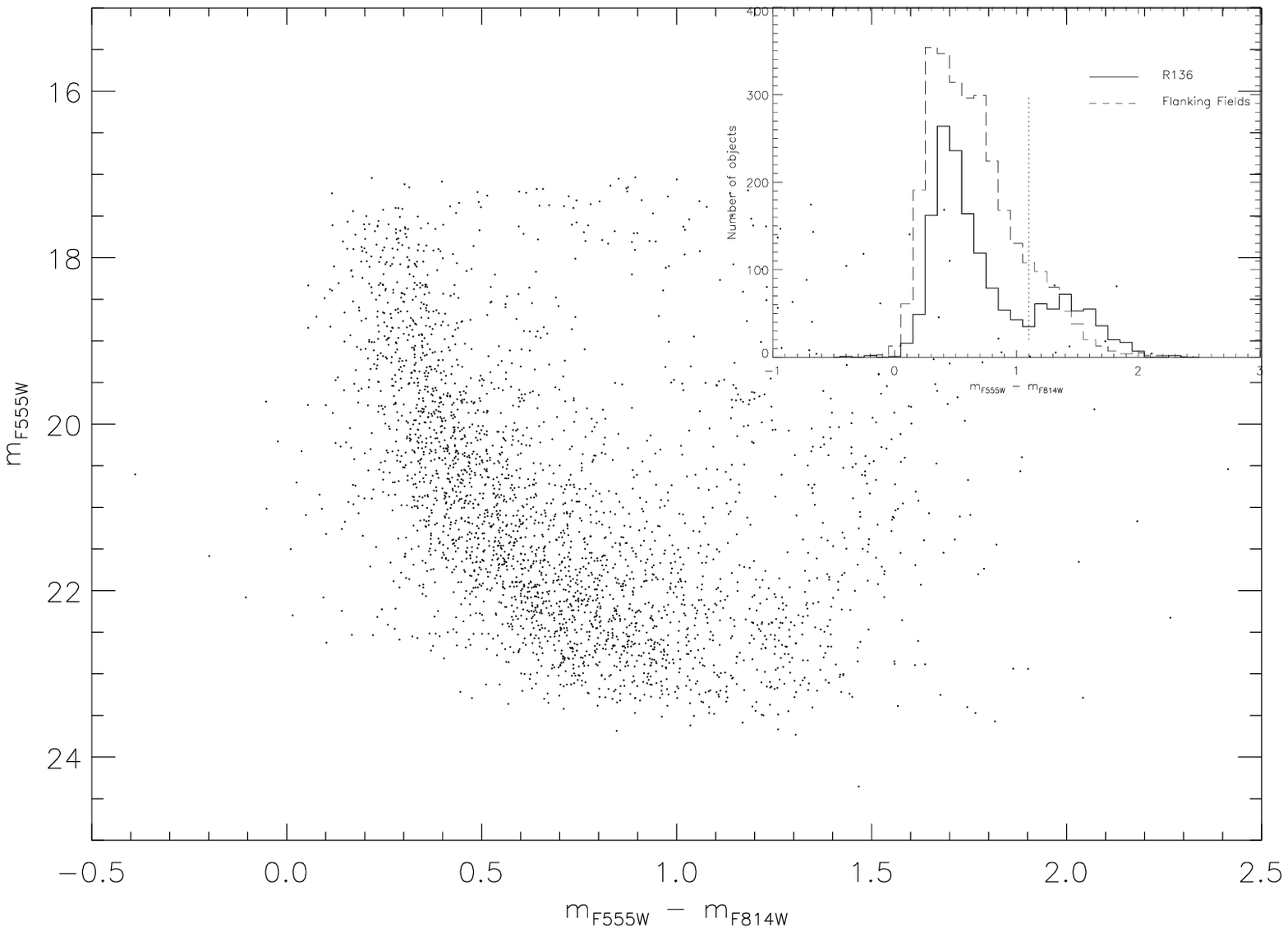}
\clearpage
\plotone{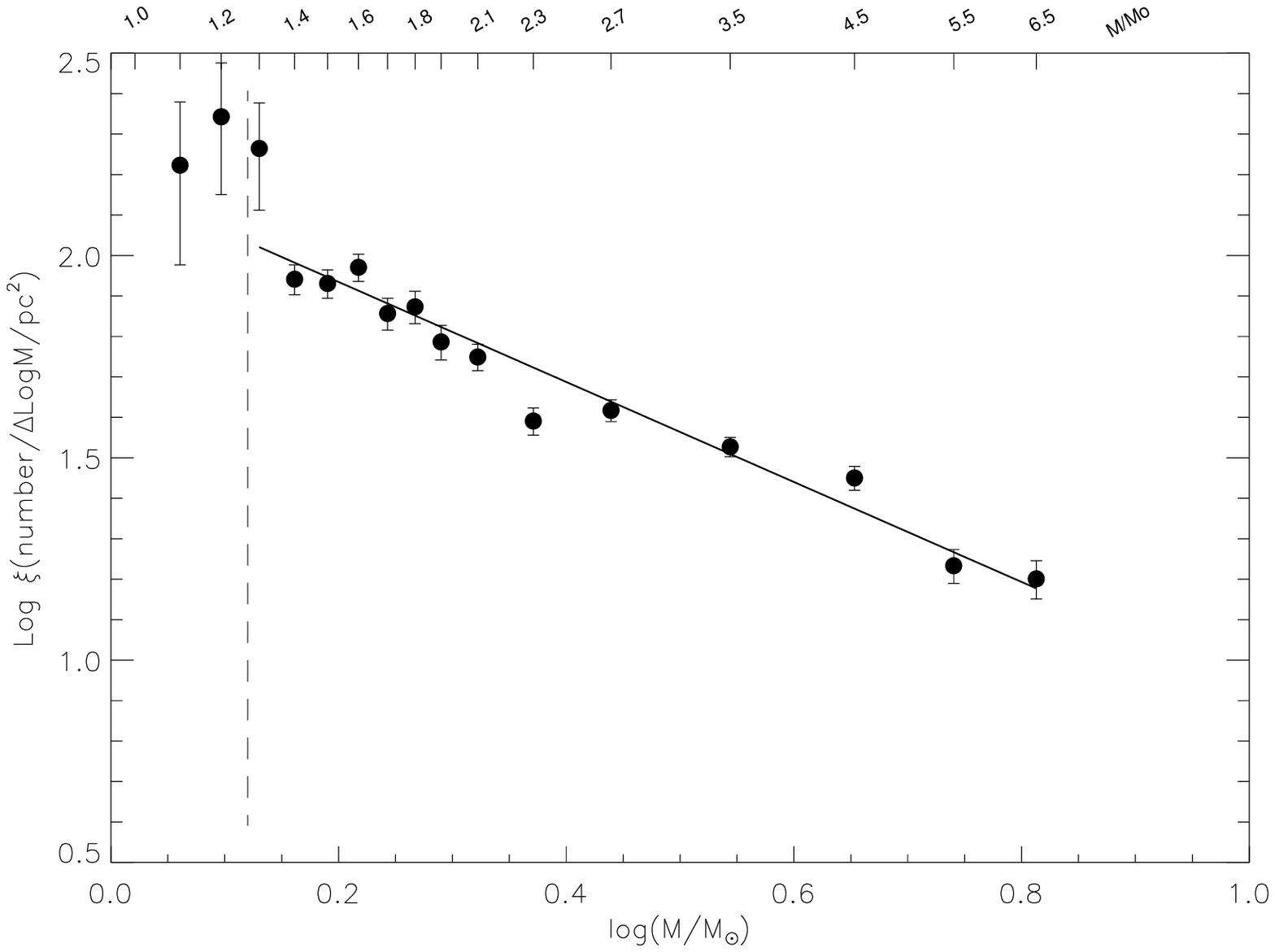}
\clearpage
\plotone{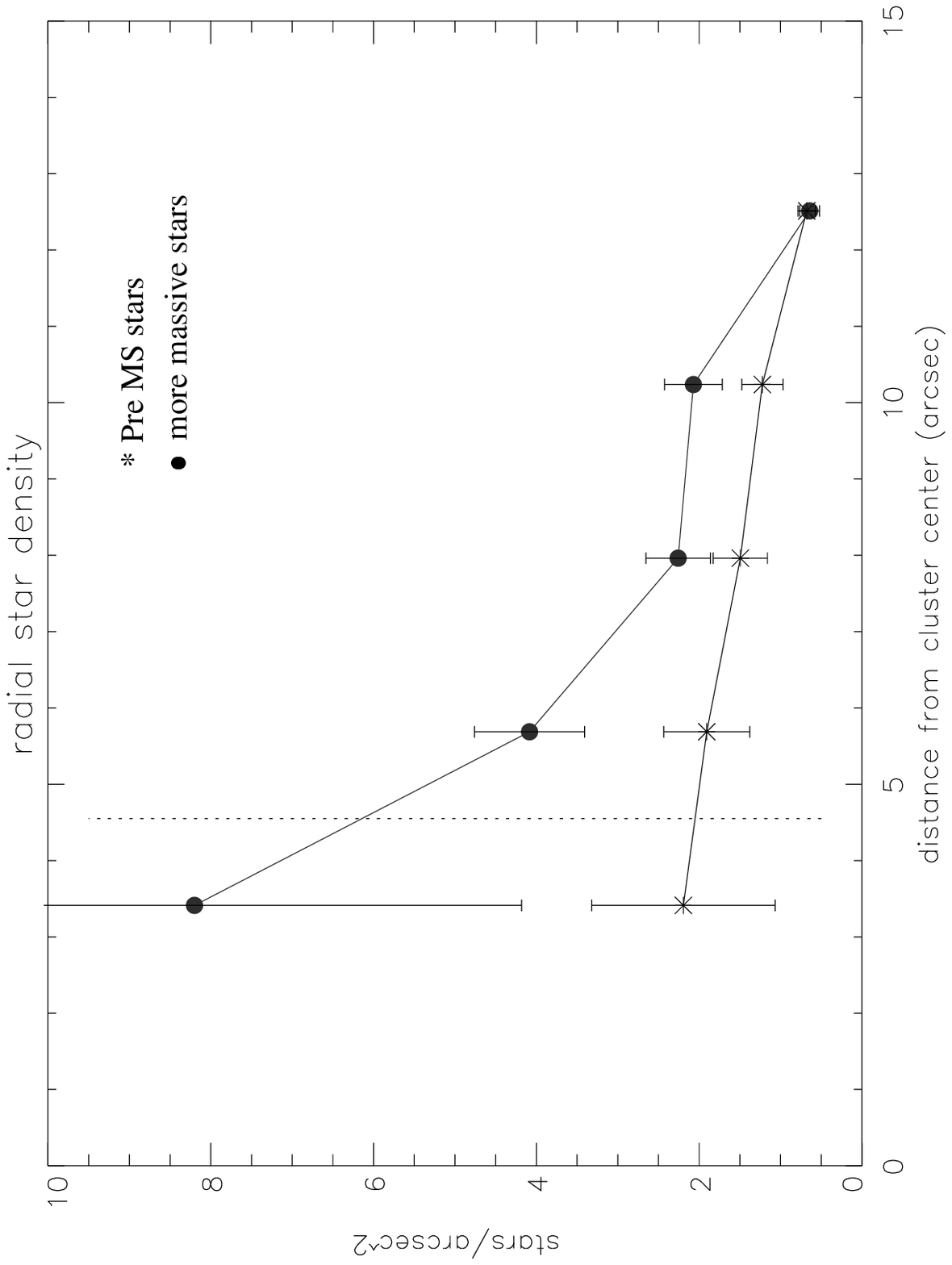}
\end{document}